\documentstyle[aps,prl,epsfig]{revtex}

\begin{document}  
\wideabs{ 
\title{Collapse dynamics of trapped Bose-Einstein Condensates}  
\author{L. Santos$^{1}$ and G. V. Shlyapnikov$^{1,2,3}$}
\address{(1)Institut f\"ur Theoretische Physik, Universit\"at Hannover,
 D-30167 Hannover,
Germany\\ 
(2) FOM Institute for Atomic and Molecular Physics,
        Kruislaan 407, 1098 SJ Amsterdam, The Netherlands\\
(3) Russian Research Center Kurchatov Institute, Kurchatov Square, 123182
 Moscow, Russia\\  
}
\maketitle  
  
\begin{abstract}
We analyze the implosion and subsequent explosion of a trapped condensate 
after the scattering length is switched to a negative value. 
Our results compare very well qualitatively and fairly well quantitatively 
with the results of recent experiments at JILA.
\end{abstract}  
}



Bose-Einstein condensates (BECs) in trapped ultracold 
atomic gases \cite{BEC,Li,Rb85,C} are strongly influenced by the atom-atom
interactions.  These interactions are characterized by a single parameter, the
$s$-wave scattering length $a$.  For $a>0$ the interatomic potential is
repulsive and the condensate is stable.  On the other hand, if $a<0$ the
interaction is attractive and a uniform condensate is unstable against local
collapses.  The trapping potential stabilizes a condensate with a sufficiently
small  number of particles $N<N_c$, where $N_c$ is a critical value below
which the  spacing between the trap levels exceeds the attractive mean-field
interparticle interaction \cite{review}. Trapped condensates with $a<0$ have 
been obtained in experiments with $^7$Li at Rice \cite{Li}, and recently at ENS
\cite{C}.


Over the last decade, the creation of condensates with tunable interparticle
interactions (tunable BEC) has attracted a great deal of interest. There have
been several proposals on how to modify the scattering length $a$
\cite{Verhaar,Fedichev96,You}. The idea of varying the magnetic field and
meeting Feshbach resonances \cite{Verhaar} has been successfully implemented 
for Na condensates at MIT \cite{FeshbachNa}.  However, the Na
experiment was limited by large inelastic losses \cite{Stenger}. Recently, a
tunable BEC of $^{85}$Rb atoms has been realized at JILA, without large
particle losses \cite{Rb85,Donley01}. These experiments constitute an
excellent tool for analyzing the influence of interatomic interactions on
the BEC properties. 


The JILA experiment \cite{Donley01} allows for the creation of large
condensates with $a=a_{init}\ge 0$ and a subsequent sudden change of the
scattering length  to $a=a_{collapse}<0$.  After this change of $a$,
the condensate undergoes an implosion (collapse) followed by the ejection of
relatively hot atoms (burst atoms), which form a halo surrounding a core of
atoms at the trap center (remnant atoms). 


The collapse of a spatially homogeneous condensate is described by the
well-known nonlinear Schr\"odinger equation (NLSE) which in the context of
BECs is called the Gross-Pitaevskii (GP) equation. This
collapse has a variety of analogies, such as self-focusing of wave beams 
in nonlinear media, collapse of Langmuir waves,
etc. (see e.g. \cite{Zakharov,Vlasov,Kosmatov} and refs.~therein). The 
collapse of  the solutions of NLSE has been extensively investigated  
and it has been found that the dimensionality of the system plays a crucial 
role \cite{Zakharov,Kosmatov}. In 3D one has a weak
collapse where the singularity is reached at a finite time.
Before this happens the collapsing cloud is described by a universal Zakharov 
solution \cite{Zu}. This solution consists of quasistationary tails 
and a collapsing 
central part. The number of particles in the collapsing part continuously 
decreases, whereas the density increases.
The dynamics of collapse in the presence of dissipation has been also 
analyzed (see \cite{Vlasov,Kosmatov} and refs.~therein).  The dissipation 
is introduced through a nonlinear imaginary (damping) term in the NLSE,  
which prevents the appearance of the singularity if the nonlinearity is at
least quintic.  

The collapse of a trapped condensate has been recently analyzed in several
theoretical papers \cite{Kagan,Kagan98,Eleftheriou,Saito,Berge,Saitonew}.
Kagan {\it et al} \cite{Kagan98} argued that three-body recombination 
should be explicitly included in the GP equation as an imaginary loss term.
The recombination "burns" only part of the condensed atoms and prevents a 
further collapse of the cloud once the peak density becomes
so high that the three-body loss rate is already comparable with the
mean-field interaction. Then the collapse turns to expansion and the
trapped condensate can undergo macroscopic oscillations accompanied by
particle losses. Kagan {\it et al} considered the case of a comparatively
large recombination rate constant, where a single collapse does not have an
internal structure. By using the formalism of Ref. \cite{Kagan98}, Saito and 
Ueda \cite{Saito} have observed  rapid
intermittent implosions of the collapsing BEC cloud.  This resembles the
distributed collapse discussed by Vlasov {\it et al} \cite{Vlasov} in the
context of collapsing cavities in plasmas. Saito and Ueda estimated the
energy  of the atoms released during the explosion, and predicted the
formation of  nonlinear patterns in the course of the collapse
\cite{Saitonew}.  

A different approach was suggested by Duine and
Stoof \cite{Duine}, who proposed binary collisions as the source of burst
atoms in the JILA experiments. Finally, K\"ohler and Burnett \cite{Kohler2}
have recently analyzed the JILA collapsing condensates with the help of a
non-Markovian nonlinear Schr\"odinger equation, suggesting that the burst
atoms can be formed due to the violation of the common $s$-wave
scattering approximation.


In the present paper, we analyze the implosion and subsequent explosion of the 
BEC in the conditions of the recent experiments with $^{85}$Rb at JILA
\cite{Donley01}. A detailed analysis of measurable quantities is
provided by numerical simulations of the GP equation which includes
three-body recombination losses as proposed in Ref. \cite{Kagan98}. Our results
agree fairly well with the data of JILA.


We consider a Bose-Einstein condensate of initially $N$ atoms of mass $m$ 
confined in a cylindrically symmetric harmonic trap. 
We restrict ourselves to the trap employed in $^{85}$Rb experiments, 
i.e. a cigar-shaped trap with axial frequency $\omega_z=2\pi\times 6.8$Hz, 
and radial frequency $\omega_\rho=2\pi\times 17.5$Hz \cite{Rb85,Donley01}. 
Assuming a sufficiently low temperature and omitting the presence of an
initial thermal cloud, the behavior of the condensate wavefunction $\psi$ is
governed by the GP equation (cf. \cite{Kagan98}):
\begin{equation} i\hbar\dot\psi= \left (- \frac{\hbar^2\nabla^2}{2m} +  V({\bf
r})+g|\psi|^2- i\frac{\hbar L_3}{12} |\psi|^4 \right )\psi,
\label{GPE}
\end{equation}
where $V({\bf r})=m(\omega_\rho^2 \rho^2 + \omega_z^2 z^2)/2$ is the trapping 
potential, $g=4\pi\hbar^2 a/m$, and $\psi$ is normalized to $N$.
The last term in the rhs of Eq.(\ref{GPE}) describes three-body recombination 
losses. The quantity $L_3$ is the recombination rate constant for an ultra-cold 
thermal cloud, and the numerical factor $1/12$ accounts for the reduction of 
the recombination rate by a factor of 6 in the condensate.

As in the conditions of the JILA experiment, we consider an initial scattering
length $a_{init}\geq 0$. For this value of $a$ we obtain the ground-state
condensate wavefunction  by evolving Eq.\ (\ref{GPE}) in imaginary time. At
$t=0$ the scattering length is abruptly switched to a value $a_{collapse}<0$.
The simulation of the subsequent dynamics under the conditions of JILA involves a very 
demanding numerical procedure, due to very different time and 
distance scales at $t=0$, during the collapse, and after the explosion. 
In our simulation we have numerically solved Eq.(\ref{GPE}) 
by means of the Crank-Nicholson algorithm with variable spatial and time
steps. We have taken a special care of the time and spatial 
numerical discretizations and checked that our results do not change
significantly when a more accurate sampling is used. 

An additional problem in simulating the BEC dynamics from Eq.(\ref{GPE}) is the 
lack of knowledge of exact values of the
three-body rate constant $L_3$. The experiments \cite{Roberts} based on
the measurement of losses in thermal clouds set an upper bound for
$L_3$, due to difficulties to distinguish between two- and 
three-body losses. The rate constant $L_3$ is safely determined only
far from the Feshbach resonance ($154.9$G for $^{85}$Rb).  A value
of  $L_3=4.24\times 10^{-24}$cm$^6/s$ has been measured at $250$G where 
$a=-336a_0$ \cite{Roberts}. 
The existing predictions for $L_3$ are related to the case of large
positive $a$ \cite{Fedichev,Braaten1}. For the recombination to deeply
bound states, which is the case at $a<0$, the predictions contain a number of
phenomenological parameters \cite{Braaten}. In our simulations we rely on
the experimental value of $L_3$ at $250$G and assume the dependence
$L_3\propto a^2$ to obtain $L_3$ for magnetic fields in which $|a|$ is
smaller ($a<0$). This particular choice is within the error bars of the
JILA measurements of $L_3$ \cite{Roberts} and leads to a fairly good agreement
between our calculations and the JILA experimental results \cite{Donley01}
for the dynamics of collapsing condensates. 

As discussed above, the condensate collapses if $N$ is larger than a critical
value. In the absence of three-body losses, the cloud collapses
continuously and the central density approaches infinity at a 
finite time $t_{collapse}$. After some time from the start of the collapse, the
cloud becomes spherical and is described by the universal Zakharov solution 
\cite{Zu}. The size of the central part of the cloud decreases
as $(t_{collapse}-t)^{1/2}$ and the central density increases as
$1/(t_{collapse}-t)$. In this stage of the collapse the presence of the
trapping potential is not important (see \cite{Berge}). We have tested the
appearance of the universal Zakharov solution in our simulations. 

The presence of three-body recombination changes the situation drastically.
Once the central density becomes such that $g|\psi(0,t)|^2 \simeq
(\hbar L_3/12)|\psi(0,t)|^4$, the collapse stops since the recombination 
losses prevent further increase of the central density \cite{Kagan98}. 
We have checked that for most of realistic values of $L_3$ the Zakharov
solution is not realized in the course of the contraction to maximum density.
For comparatively small $L_3$ the maximum
density of the cloud is rather high, and the number of particles in the
central part of the cloud is small. These particles are rapidly burned by the
recombination and the central density drops. However, the central region is
then quickly refilled by the flux of particles from the wings of the spatial
distribution. Therefore one obtains a set of intermittent collapses, i.e. the 
collapse becomes distributed \cite{Vlasov,Saito}.
As each intermittent collapse burns only a very small number of atoms the 
total number of particles presents a smooth time dependence.

In our calculations we have analyzed the implosion 
and successive 
explosion of a condensate for the initial number of atoms $N=6000$
and $N=15000$, and for $a_{collapse}$ ranging from $-25a_0$ to $-300a_0$.
We have considered $a_{init}=0$ for the case of $N=6000$, and  $a_{init}=7a_0$
for $N=15000$, in order to compare our results with those at JILA 
\cite{Donley01}. Typically, we observe that the condensate contracts
mostly radially and reaches a maximum central density after a time
$t_{collapse}$ which ranges from $0.5$ ms to several ms. Then intermittent 
collapses occur. Close to the maximum density in each intermittent 
collapse, the central region of the collapsing  condensate
becomes spherical. As expected, the collapse stops when  $(\hbar L_3/12)
|\psi(0,t)|^4 \simeq g|\psi(0,t)|^2$ \cite{Kagan}.  At this maximum density
the total number of condensed atoms ($N_{tot}$) decreases due to  
recombination losses. Due to the presence of a set of intermittent collapses,
the time dependence of $N_{tot}$ shows a step-wise decay. However, the
average over short time intervals of the order of the time interval between
neighbouring intermittent collapses allows us to fit $N_{tot}$ by an
exponential $\exp(-t/\tau_{decay})$. After the BEC explodes, we observe the
formation of a dilute halo of burst atoms surrounding a central cloud of
atoms. This reproduces qualitatively the picture observed at JILA. 

\begin{figure}[ht] 
\begin{center}\ 
\epsfxsize=4.70cm 
\hspace{0mm} 
\psfig{file=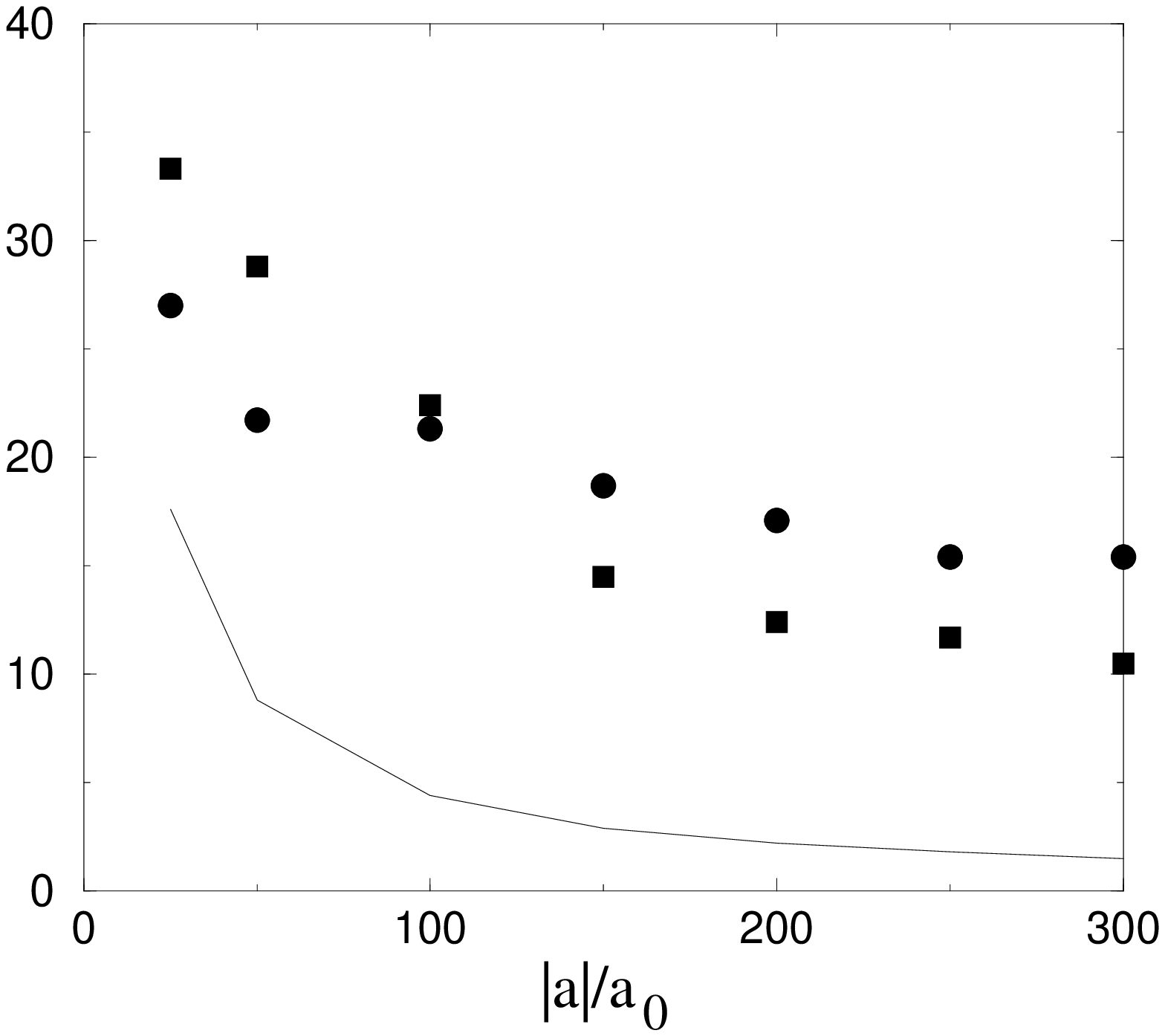,width=4.7cm}\\[0.1cm]
\end{center} 
\caption{The fractions $N_{burst}/N$ (circles) and $N_{remnant}/N$ (squares)
in $\%$ versus $a_{collapse}$ for $N=6000$, $a_{init}=0$. The solid curve shows
$N_{cr}/N$.} \label{fig:1}  
\end{figure}
\begin{figure}[ht] 
\begin{center}\ 
\epsfxsize=4.7cm 
\hspace{0mm} 
\psfig{file=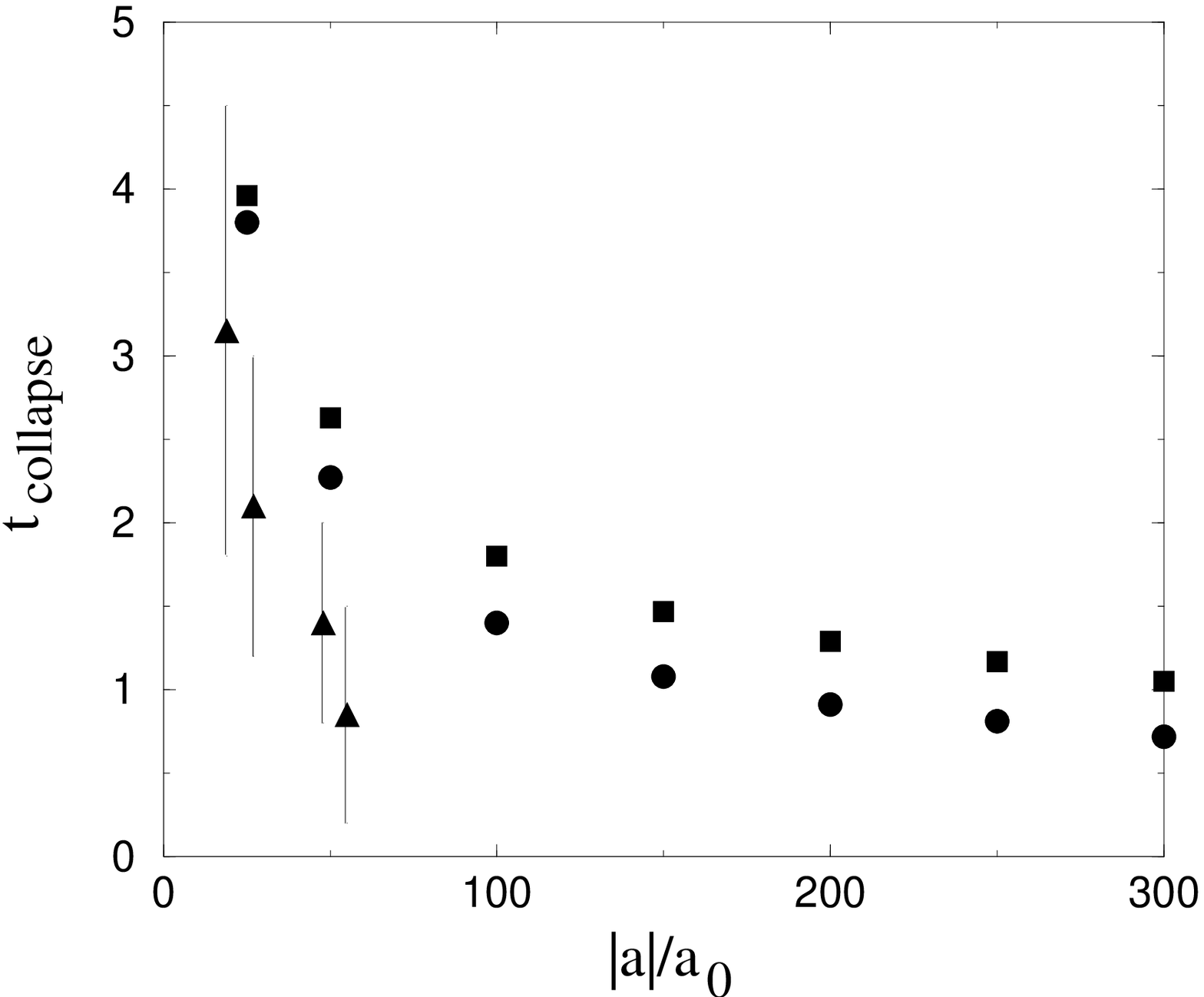,width=4.7cm}\\[0.1cm]
\psfig{file=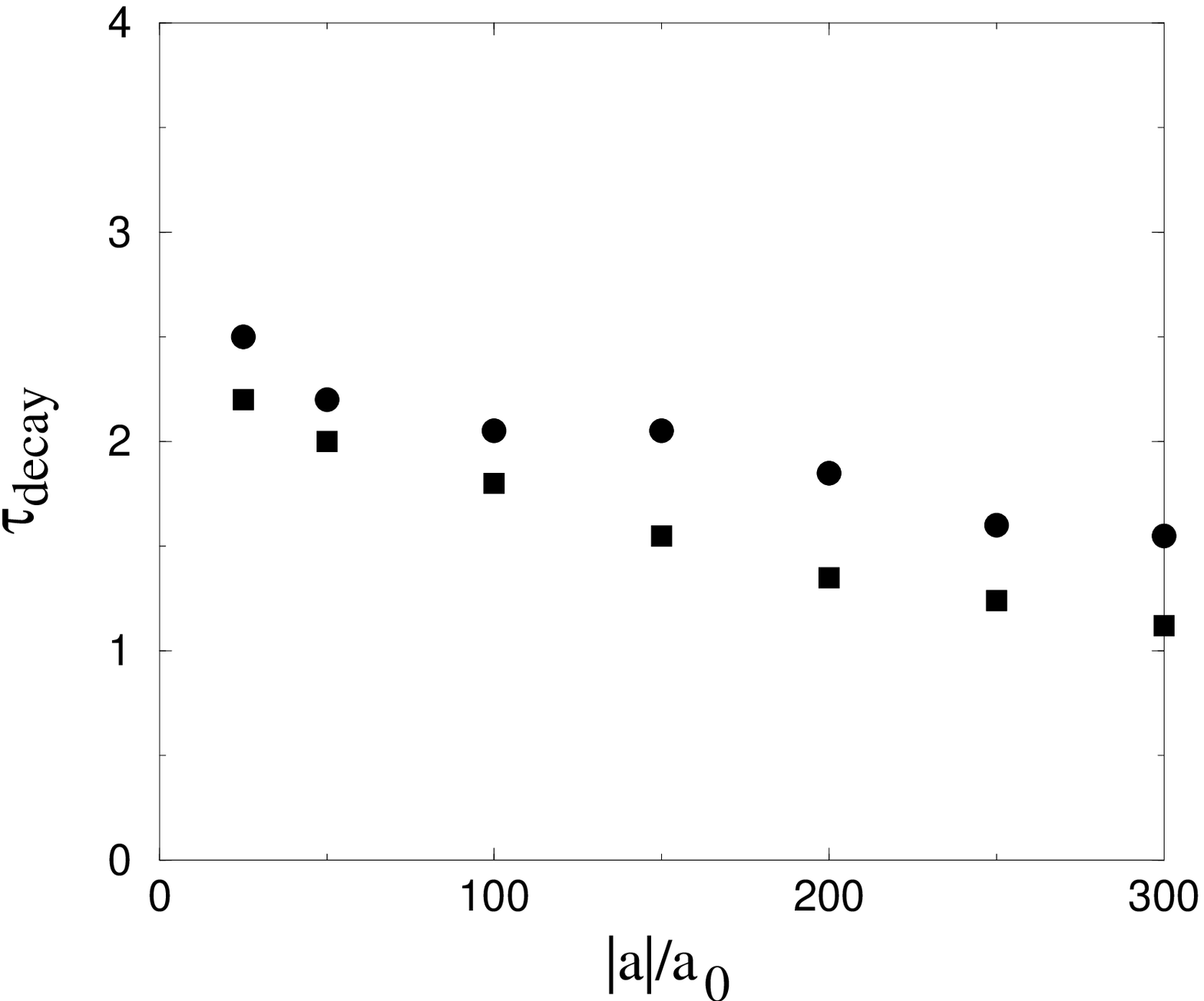,width=4.7cm}\\[0.1cm]
\end{center}
\caption{The times $t_{collapse}$ (upper figure) and $\tau_{decay}$ (bottom
figure) versus $a_{collapse}$ for $N=6000$, $a_{init}=0$ (circles),
and $N=15000$, $a_{init}=7a_0$ (squares). In the upper figure the
triangles show the experimental results of JILA for $N=6000$.}
\label{fig:2}  
\end{figure}

We calculate various quantities: $t_{collapse}$, $\tau_{decay}$, 
the number of burst ($N_{burst}$) and remnant ($N_{remnant}$) atoms, 
and the axial and radial energy of the burst atoms. 
We determine the number of burst atoms by
integrating the condensate density over the axial coordinate and fitting
the tail of the obtained radial profile by a Gaussian. 
The axial and radial energies per particle in the burst were calculated
by averaging the axial and radial Hamiltonians, 
$H_z=-\hbar^2\nabla^2/2m+m\omega_z^2 z^2/2$ and  
$H_\rho=-\hbar^2\nabla^2/2m+m\omega_{\rho}^2\rho^2/2$,
over the density distribution. Since the presence of the
remnant cloud may introduce errors in the determination of the burst energies,
we have excluded the central region from the average of $H_z$ and $H_\rho$. We
prevent possible errors by extracting central regions of different widths
ranging from one to several harmonic oscillator lengths. After a typical
simulation time of $20$ ms, we have checked that our results for the burst
energy per particle are independent of the width of the
excluded central region.


In Fig.\ \ref{fig:1} we present the fractions $N_{burst}/N$ and
$N_{remnant}/N$ versus $a_{collapse}$ for $N=6000$ at a time of $20$ ms. 
After this time $N_{burst}$ and $N_{remnant}$ reach stationary values. 
In the same figure we depict the critical value $N_{cr}=k a_{ho}/|a_{collapse}|$, 
were $k=0.46$ is the stability coefficient for $^{85}$Rb \cite{Roberts01}, 
and $a_{ho}=\sqrt{\hbar/m\bar\omega}$ with $\bar\omega=(\omega_{\rho}^2\omega_z)^{1/3}$. 
A similar picture has been obtained for $N=15000$. The burst fraction 
$N_{burst}/N$ varies between $15\%$ and $25\%$ and only weakly depends on $N$
and $a_{collapse}$. This is in good agreement with the 
results of JILA \cite{Donley01}, where $N_{burst}/N\simeq 20\%$. One can also
see that $N_{remnant}>N_{cr}$, which is expected and is in agreement with  
the experiments at JILA.

Fig.\ \ref{fig:2} displays the dependence of $t_{collapse}$ and
$\tau_{decay}$ on $a_{collapse}$ for $N=6000$ and $N=15000$, respectively. 
As observed, neither characteristic time changes significantly with
changing $N$. The time of collapse ranges from $0.5$ to $4$ ms
for considered values of $a_{collapse}$. The decay time $\tau_{decay}$ 
weakly depends on $a_{collapse}$.  For $N=6000$ it ranges 
from $2.5$ ms at $a_{collapse}=-25a_0$ to $1.6$ ms at $a_{collapse}=-300a_0$.
For the same values of $a_{collapse}$ at $N=15000$, the time $\tau_{decay}$
ranges from $2.2$ to $1.1$ ms. These results are in excellent agreement with
Ref.\ \cite{Donley01}.

\begin{figure}[ht] 
\begin{center}\ 
\epsfxsize=4.7cm 
\hspace{0mm} 
\psfig{file=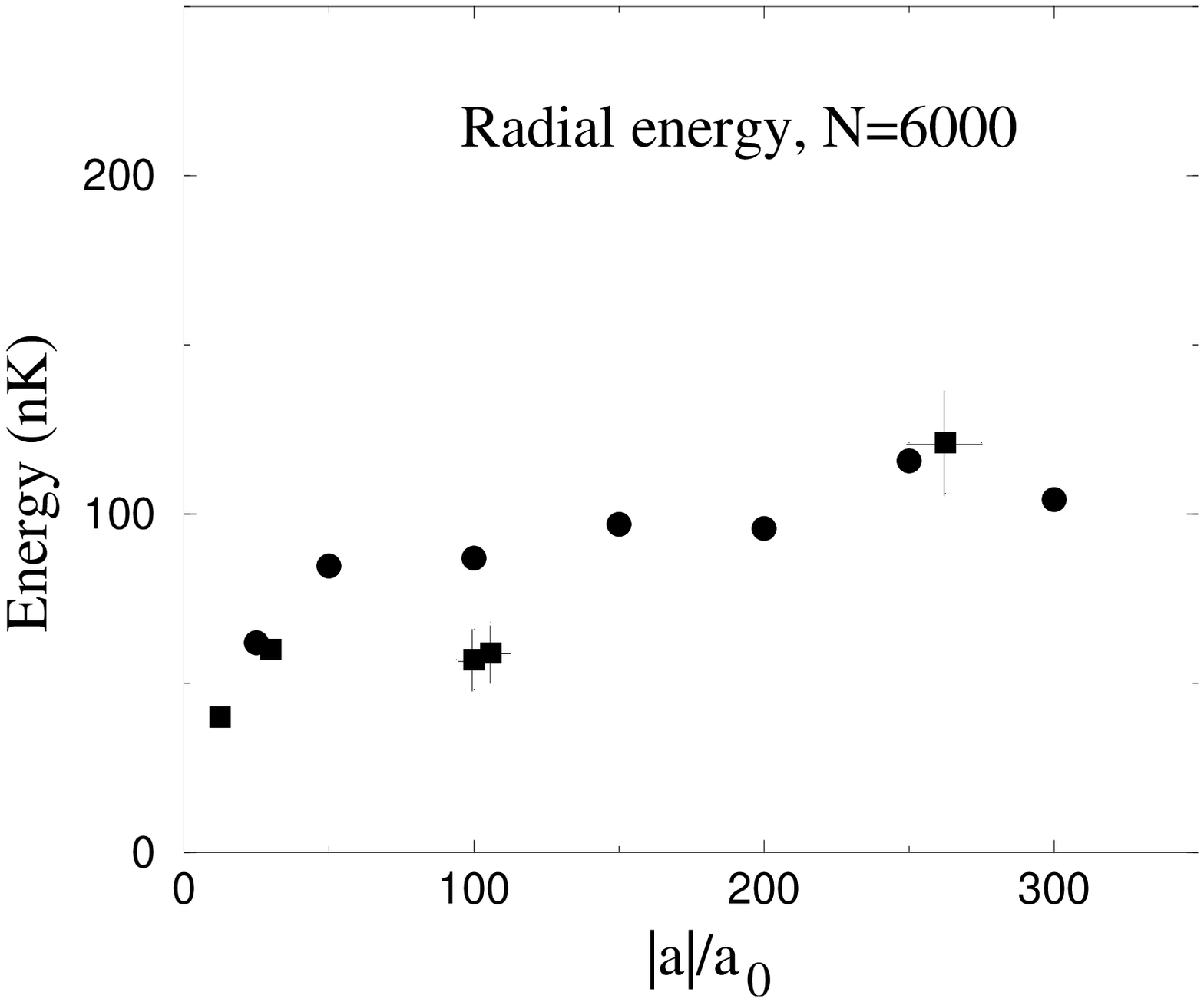,width=4.7cm}\\[0.1cm]
\psfig{file=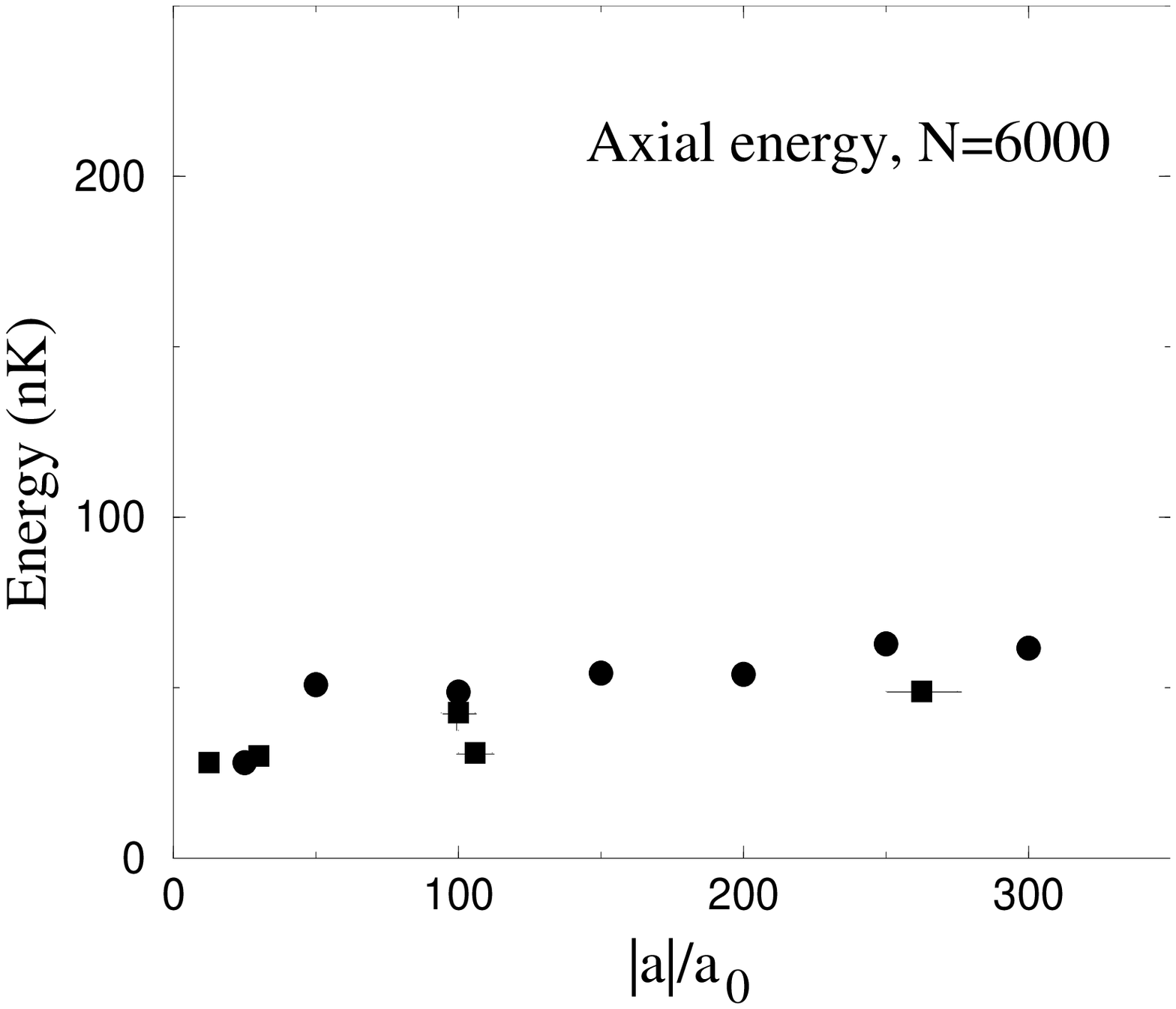,width=4.7cm}\\[0.1cm]
\psfig{file=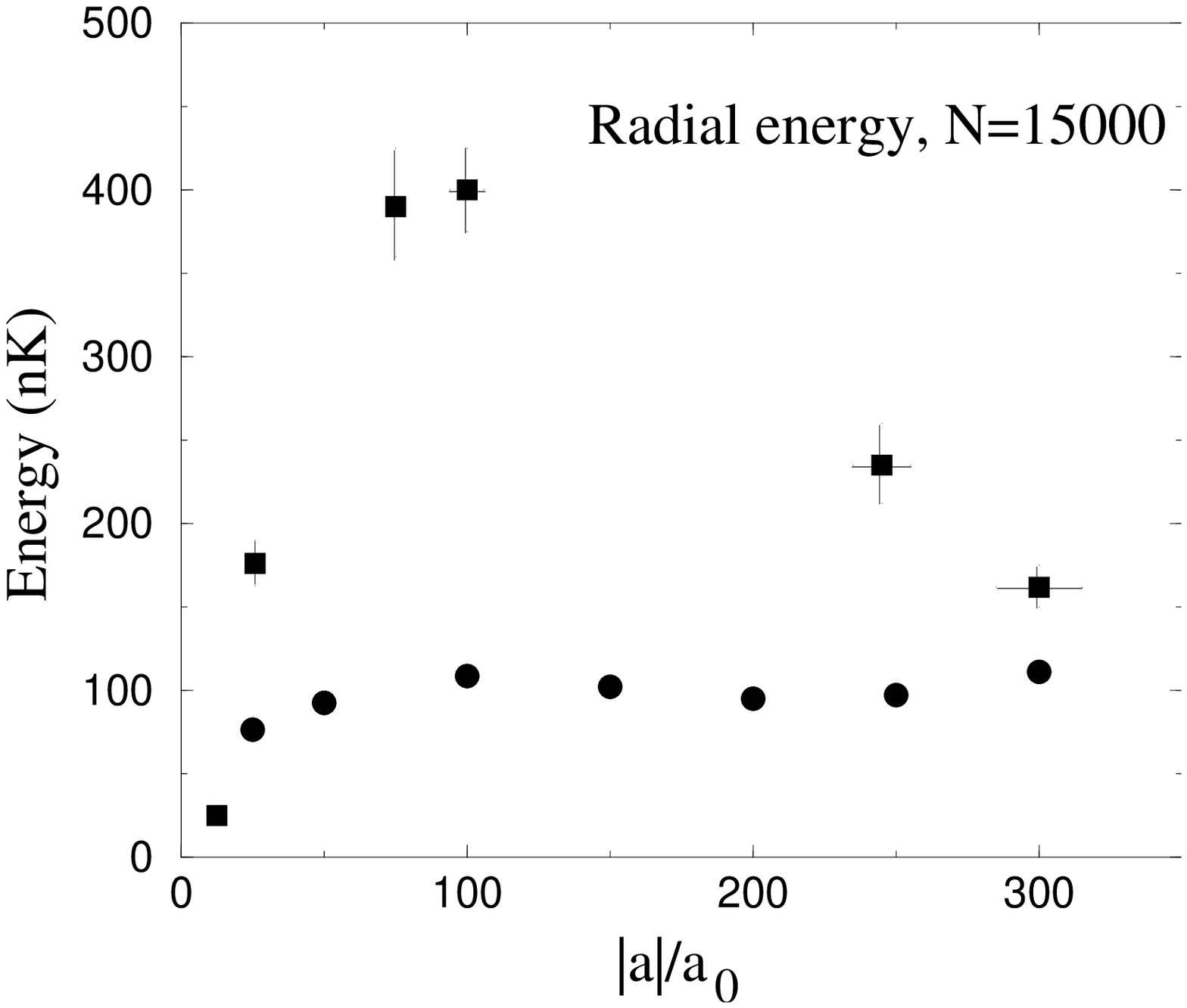,width=4.7cm}\\[0.1cm]
\psfig{file=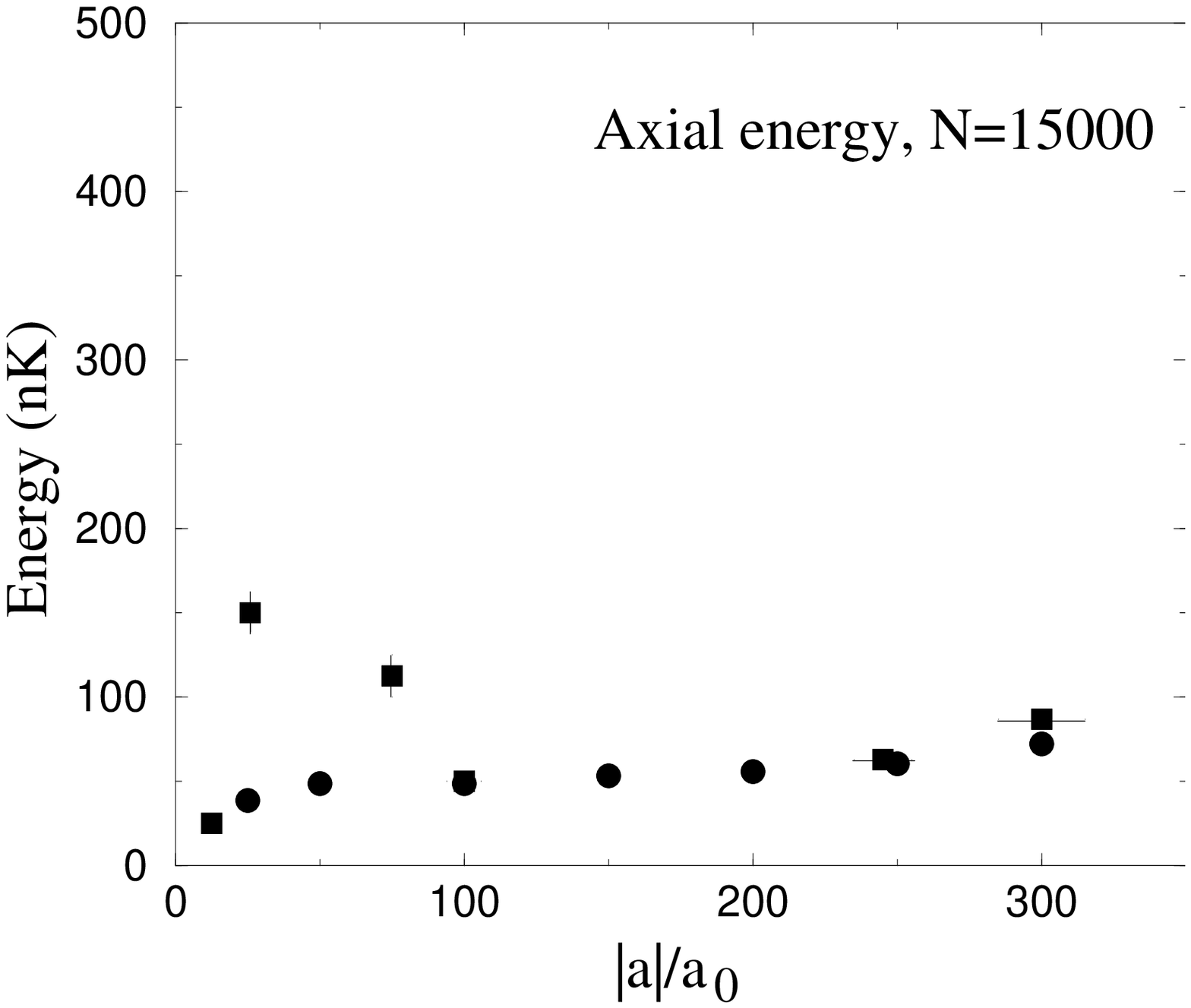,width=4.7cm}\\[0.1cm]
\end{center} 
\caption{Radial and axial energies versus $a_{collapse}$ for
$N=6000$, $a_{init}=0$, 
and $N=15000$, $a_{init}=7a_0$.
Our numerical results (circles) are compared with the experimental data 
at JILA (squares with lines indicating the error bars.}
\label{fig:3}  
\end{figure}

Fig.\ \ref{fig:3} shows the radial and axial burst energies versus
$a_{collapse}$ for $N=6000$ and $N=15000$. 
Both energies smoothly depend on $a_{collapse}$ and $N$. The radial energy is
of the order of $100$ nK, and the axial one of the order of $50$ nK.  
We have performed simulations for a large range of
values of $a$ and $L_3$ and found that the  energy of the burst atoms is
proportional to $a^2/L_3$, as observed by Saito and Ueda \cite{Saito}.  This
is expected as the burst energy should be of the order of the
maximum  mean-field interaction $g|\psi(0,t)|^2$ at the trap center for 
$t\sim t_{collapse}$, and this interaction has exactly the same scaling.

For the three-body rate constants used in our calculations, the results are 
in a fair agreement with the data of the JILA experiment. However, we are 
not able to reproduce the JILA results for the radial 
burst energy in the case of $N=15000$ and $a_{collapse}\simeq 100a_0$. 
The JILA radial burst energy for this particular set of parameters is 
by a factor of $4$ larger than that observed in our simulations, whereas 
the rest of measured energies depart by less than $50\%$ from our results.  

To summarize, we have analyzed the collapse dynamics of
trapped condensates after the scattering length is switched to a negative
value. Our analysis, based on the GP equation with 
three-body losses explicitly included, explains qualitatively and to a large
extent quantitatively the experiments performed at JILA.

We acknowledge support from Deutsche Forschungsgemeinschaft (SFB
407), the Alexander von Humboldt Foundation, the TMR Network 
'Coherent Matter Wave Interactions', the Dutch Foundations NWO and FOM, and 
from the Russian Foundation for Fundamental Research. 
L. S. wishes to thank the ZIP Program of the German Government. 
We acknowledge fruitful discussions with 
S. L. Cornish, E. A. Donley, J. L. Roberts,  
E. A. Cornell, C. E. Weiman, and M. Lewenstein.

Note added: After this work was completed we learned that 
Saito and Ueda extended their analysis, also based on the GP equation
with three-body losses, to the case of axially symmetric trap of JILA 
(second version of  \cite{Saitonew}). They calculated $t_{collapse}$ and
found the number of burst and remnant atoms as functions of the initial number
of atoms $N$ at a given value of $a_{collapse}$. Their results for
$t_{collapse}$ agree very well with ours.

\end{document}